\documentclass[11pt,a4paper,twoside]{article} 
\usepackage{amsmath}
\usepackage{amsfonts}
\usepackage[normalem]{ulem}
\usepackage{color}
\usepackage{rotating}
\usepackage{graphicx}
\usepackage[T1]{fontenc} 
\usepackage{bm}
\usepackage{hyperref}
\usepackage{version}

\numberwithin{equation}{section}

\begin{document}

\noindent
{\bf
{\Large On the introduction of a boundary in topological field theories 
}} 

\vspace{.5cm}
\hrule

\vspace{1cm}

\noindent
{\large{Andrea Amoretti$^{a,b,}$\footnote{\tt andrea.amoretti@ge.infn.it },
Alessandro Braggio$^{b,c,}$\footnote{\tt alessandro.braggio@spin.cnr.it }, 
Giacomo Caruso$^{a,b,}$\footnote{\tt giacomo.caruso@ge.infn.it }, \\
Nicola Maggiore$^{a,b,}$\footnote{\tt nicola.maggiore@ge.infn.it },
Nicodemo Magnoli$^{a,b,c,}$\footnote{\tt nicodemo.magnoli@ge.infn.it}
\\[1cm]}}

\noindent
{{}$^a$ Dipartimento di Fisica, Universit\`a di Genova,\\
via Dodecaneso 33, I-16146, Genova, Italy\\
{}$^b$ I.N.F.N. - Sezione di Genova\\
{}$^c$  CNR-SPIN, Via Dodecaneso 33, 16146, Genova, Italy}
\vspace{1cm}

{\tt Abstract~:}
We study the consequences of the presence of a boundary in topological field theories in various dimensions. We characterize, univocally and on very general grounds, the field content and the symmetries of the actions which live on the boundary. We then show that these actions are covariant, despite appearances. We show also that physically relevant theories like the 2D Luttinger liquid model, or the 4D Maxwell theory, can be seen as boundary reductions of higher dimensional topological field theories, which do not display local observables. 

\vfill\noindent
{\footnotesize 
{\tt Keywords:}
Quantum Field Theory, Duality in Gauge Field Theories, Discrete and Finite Symmetries, Gauge Symmetry, Boundary Quantum Field Theory.
\\
{\tt PACS Nos:} 
03.70.+k	Theory of quantized fields;
11.10.-z	Field theory;
}

\section{Introduction}

It is well known that topological field theories acquire local observables only once a boundary is introduced. One of the most remarkable examples is non-Abelian 3D Chern-Simons theory, whose boundary embeds all rational 2D conformal field theories \cite{Moore:1989yh}. In this paper we consider topological field theories in 3D, 4D and 5D with a planar boundary. In all cases, on the boundary non topological actions are found, displaying non trivial physical content. The aim of this paper is to give a unitary plot of the boundary reduction of topological field theories, stressing some crucial points which were not fully covered in previous papers \cite{Blasi:2010gw,Blasi:2011pf,Amoretti:2012kb,Amoretti:2013nv}, starting from the role of gauge symmetry. The boundary is realized in the action by means of a local term, proportional to the Heaviside step function. The introduction of the boundary spoils the gauge invariance of those theories whose Lagrangian transforms, under gauge transformations, into a total derivative. This property is peculiar to  topological field theories of the Schwarz type \cite{Birmingham:1991ty} considered in this paper. A boundary term compatible with locality and power counting is added. The imposition that the complete action, including the boundary term, is gauge invariant gives rise to a condition, verified on-shell, which is fundamental in order to identify the field content and the symmetries of the lower dimensional theory living on the boundary. The boundary conditions make also possible another kind of identification. It is known, for instance, that certain boundary conditions, called duality relations in \cite{Aratyn:1984jz}, lead to claim the existence of fermionic degrees of freedom on the boundary, despite the fact that the bulk actions are entirely bosonic. An explicit construction of these fermionic modes has been done in \cite{Amoretti:2013xya}. This is most relevant for the bulk 3D and 4D topological BF models, whose boundary reductions may describe the physics of topological insulators \cite{Hasan:2010xy,zhang}. Hence, the boundary degrees of freedom are direct consequences both of the gauge symmetry and of the boundary conditions on the fields. But the role of gauge symmetry is even more relevant. In fact, on the boundary a residual gauge symmetry survives. It is functionally described by Ward identities that generate relations between Green functions. These, written in terms of boundary fields, are interpreted as canonical commutation relations. The lower dimensional theory living on the boundary is obtained by requiring that its equations of motion are compatible with the commutation relations coming from the Ward identities and with the boundary conditions.
\\
{
We stress that the same starting topological model might give rise to different models of boundary dynamics, depending on the boundary terms introduced in the action. For instance, we chose to preserve Lorentz invariance on the boundary, which seems reasonable, but physical reasons to relax this choice might occurr. Even so, a kind of ambiguity remains, because of some non universal constants, which are free parameters for the theory. 
}
\\
The outline of the paper is as follows. In section \ref{secbf3} we give a detailed account of our method applied to the 3D BF model.
This leads to identify the 2D theory of Luttinger liquid \cite{Tomonaga:1950zz,Luttinger:1963zz} as the edge reduction of 3D BF theory.  In section \ref{secbf45} we extend our method to the 4D and 5D BF model. The Cho and Moore model \cite{Cho:2012tx} is recovered on the boundary of 4D BF model which on-shell reduces to a 3D Maxwell theory or a free scalar theory. On the boundary of 5D BF model we obtain an action which on-shell corresponds to the Kalb-Ramond theory \cite{Kalb:1974yc} or a free scalar theory. Finally, in section \ref{secbc} we consider a theory which we call BC model, built by means of two rank-2 tensors and no gauge fields.
Also for this model we recover the Maxwell theory as a on-shell reduction of the boundary action. For each model the boundary theory is selected by Time Reversal discrete symmetry. It is proved the covariance of the boundary theories we obtained, which is not an obvious issue, since this approach origins from the breaking of covariance by means of the introduction of a boundary and by the choice of a non-covariant gauge fixing in the bulk. Our results are summarized in section \ref{seccon}.
\vspace{0.5cm}
\hrule
\vspace{0.1cm}
\hrule
\vspace{0.5cm}
The notations used throughout the paper are (in D dimensions):
\begin{equation}
\begin{split}
\mu,\nu,... &=\{0,1,...,D-1\}
\\
i,j,...&=\{0,1,...,D-2\}
\\
\alpha,\beta,...&=\{1,...,D-2\}
\\
x=x_{\mu}&=(x_0,x_1,..., x_{D-1})
\\
X=X_{i}&=(x_0,x_1,..., x_{D-2})
\\
g_{\mu\nu}&=diag(-1,1,1,...),
\end{split}
\end{equation}
for $D=3$:
\begin{equation}
\begin{split}
\epsilon^{012}&=1
\\
\epsilon^{ij}&=\epsilon^{ij2},
\end{split}
\end{equation}
for $D=4$:
\begin{equation}
\begin{split}
\epsilon^{0123}&=1
\\
\epsilon^{ijk}&=\epsilon^{ijk3}
\\
\end{split}
\end{equation}
for $D=5$:
\begin{equation}
\begin{split}
\epsilon^{01234}&=1
\\
\epsilon^{ijkl}&=\epsilon^{ijkl4}
\\
\end{split}
\end{equation}

\vspace{0.5cm}
\hrule
\vspace{0.1cm}
\hrule
\vspace{0.5cm}


\section{The Abelian 3D BF theory with boundary}\label{secbf3}

\subsection{The action and its symmetries}

We consider the Abelian 3D BF model, defined on the flat Minkowski space-time  with a boundary on the plane $x^2=0$. The action of the model is:
\begin{equation}
\mathcal S_{bulk}=\int{d^3x\,\epsilon^{\mu\nu\rho}\left[\partial_{\mu}A_{\nu}B_{\rho} + k A_{\nu}\partial_{\mu}B_{\rho} \right] \theta(x^2)}. \label{sbf3}
\end{equation}
It describes the interaction between two gauge fields, $A_{\mu}$ and $B_{\mu}$, and the presence of the boundary is implemented by the introduction of the Heaviside step function $\theta(x^2)$. {Since we are in the Abelian case, the canonical mass dimensions of the fields are not fixed, the only condition on them is $\left[A\right] +\left[B\right]=2$. We make the choice:}
\begin{equation}
\left[A\right]=\left[B\right]=1.
\end{equation}
The action \eqref{sbf3} is invariant under the discrete Time Reversal symmetry $T$ defined as: 
\begin{align}
&Tx^0=-x^0\notag
&
&Tx^{\alpha}=x^{\alpha}
\\
&TA_0=A_0
&
&TB_0=-B_0 \label{tr3}
\\
&TA_{\alpha}=-A_{\alpha}\notag
&
&TB_{\alpha}=B_{\alpha},
\end{align}
{under which the field $A_\mu$ transforms as the electromagnetic potential, while $B_\mu$ can be viewed as a kind of spin current.} This reflects the possibility to identify the 3D BF theory as the model for the 2D topological insulators \cite{Cho:2012tx}.
\\
The presence of the Heaviside step function modifies the usual rule for integration by parts into:
\begin{equation}\label{parts3}
\int{d^3x\, \epsilon^{\mu\nu\rho}\partial_{\mu}A_{\nu}B_{\rho}\theta(x^2)} + 
\int{d^3x\, \epsilon^{\mu\nu\rho}A_{\nu}\partial_{\mu}B_{\rho}\theta(x^2)} 
 + \int{d^3x\, \epsilon^{ij}A_i B_j \delta(x^2)} = 0, 
\end{equation}
where the third term comes from the differentiation of the $\theta$-function. Only two of the three terms in \eqref{parts3} are independent from each other so we can study the action in the form \eqref{sbf3}, where $k$ is a coupling constant which cannot be absorbed by field redefinitions and having in mind that the boundary term $\int{d^3x\, \epsilon^{ij}A_i B_j \delta(x^2)}$ can be obtained from the bulk ones after an integration by parts. Identity \eqref{parts3} imposes a condition on the coupling constant:
\begin{equation}
k \ne 1,
\end{equation}
otherwise the action \eqref{sbf3} would reduce to a pure boundary term. The presence of the boundary allows us to add to the action \eqref{sbf3} the most general boundary term compatible with locality and power counting:
\begin{equation}
\mathcal S_{bd}=\int{d^3x\, \left[\frac{a_1}{2}A_i A^i + \frac{a_2}{2}B_i B^i + a_3 A_i B^i \right] \delta(x^2)},\label{sbd3}
\end{equation}
where  $a_1,a_2,a_3$ are dimensionless constant parameters. The boundary action \eqref{sbd3} breaks the symmetry under $T$ (which is preserved only if $a_3=0$). We can define the action:
\begin{equation}
\mathcal S_{BF3}=\mathcal S_{bulk} + \mathcal S_{bd}\label{sb+b}
\end{equation}
The bulk equations of motion derived from \eqref{sbf3} are:
\begin{gather}\label{eom31}
\left[(1-k)\epsilon^{ij}(\partial_j B_2 - \partial_2 B_j)  \right]\theta(x^2) =0
\\
\left[(1-k)\epsilon^{ij}\partial_i B_j \right]\theta(x_2)=0\label{eom32}
\\
\left[(1-k)\epsilon^{ij}(\partial_j A_2 - \partial_2 A_j) \right]\theta(x^2) =0
\\
\left[(1-k)\epsilon^{ij}\partial_i A_j \right]\theta(x^2)=0,\label{eom34}
\end{gather}
and the boundary conditions {are obtained putting equal to zero the $\delta$-dependent part of the equations of motion deriving from the whole action \eqref{sb+b}}:
\begin{gather}
\left.a_1 A^i  + a_3 B^i -\epsilon^{ij}B_j\right|_{x^2=0}=0 \label{be31}
\\
\left.a_2 B^i + a_3 A^i + k\epsilon^{ij}A_j\right|_{x^2=0}=0.\label{be32}
\end{gather}
Without the boundary, the bulk action of the 3D BF model would be invariant under the following gauge transformations:
\begin{equation}\label{g31}
\begin{split}
\delta^{(1)}A_{\mu}&=\partial_{\mu}\varphi
\\
\delta^{(1)}B_{\mu}&=0
\end{split}
\end{equation}
and
\begin{equation}\label{g32}
\begin{split}
\delta^{(2)}A_{\mu}&=0
\\
\delta^{(2)}B_{\mu}&=\partial_{\mu}\xi,
\end{split}
\end{equation}
where $\varphi(x)$ and $\xi(x)$ are local gauge parameters. The presence of the boundary breaks the gauge invariance:
\begin{gather}
\delta^{(1)} \mathcal S_{BF3}=\int{d^2x\,\varphi\partial_i \left[k\epsilon^{ij} B_j - a_1A^i  - a_3  B^i\right]\delta(x^2)}\label{gi31}
\\
\delta^{(2)}\mathcal S_{BF3}=-\int{d^2x\,\xi \partial_i\left[ \epsilon^{ij}A_j + a_2 B^i + a_3 A^i\right]\delta(x^2)}.\label{gi32}
\end{gather}
But the gauge invariance can be restored on-shell: in fact, substituting the boundary conditions \eqref{be31} and \eqref{be32} respectively into \eqref{gi31} and \eqref{gi32}, we get:
\begin{gather}
\delta^{(1)} \mathcal S'_{BF3}=(k-1)\int{d^3x\,\varphi \epsilon^{ij}\partial_i B_j }\delta(x^2)
\\
\delta^{(2)}\mathcal S'_{BF3}=(k-1)\int{d^3x\,\xi \epsilon^{ij}\partial_i A_j }\delta(x^2).
\end{gather}
So the conditions of gauge invariance (reminding that $k \ne 1$) are:
\begin{gather}
\left.\epsilon^{ij}\partial_i B_j\right|_{x^2=0}=0\label{gic31}
\\
\left.\epsilon^{ij}\partial_i A_j \right|_{x^2=0}=0,\label{gic32}
\end{gather}
which are immediately verified on-shell since they respectively correspond to the equations of motions \eqref{eom32} and \eqref{eom34}. We will also see that these conditions are crucial in order to identify the nature of the fields on the boundary.
\\
The introduction of the boundary term \eqref{sbd3} preserves the gauge invariance of the theory and it is crucial to give non trivial dynamics on the boundary. Indeed, in the absence of the boundary term \eqref{sbd3}, the boundary conditions would lead to the constraints:
\begin{equation}
\left.A_i\right|_{x^2=0}=\left.B_i\right|_{x^2=0}=0,
\end{equation}
which would completely trivialize the boundary dynamics.

\subsection{Solutions of the boundary conditions}\label{boco}

In the previous section we have presented the model and its symmetries. Before the complete treatment of the boundary dynamics, it is important to find out which values of the constant parameters $a_1, a_2, a_3$ are solutions of the  boundary equations \eqref{be31} and \eqref{be32}.
There are two solutions which yield non trivial boundary dynamics:
\begin{enumerate}
\item
\begin{equation}
a_1 a_2 = -k, \hspace{1cm} a_1 \ne 0, \hspace{1cm} a_3=0,
\end{equation}
whose boundary conditions are combined into a unique one:
\begin{equation}\label{bcf3}
a_1 A^i-\epsilon^{ij}B_j=0.
\end{equation}
This is the only solution which extends the Time Reversal symmetry \eqref{tr3} also to the boundary.

\item
\begin{equation}\label{bbcc}
k=-1, \hspace{1cm} a_1=a_2=0, \hspace{1cm} a_3=\pm 1,
\end{equation}
whose boundary conditions are:
\begin{equation}\label{bc333}
\epsilon^{ij}A_j=\pm A^i, \hspace{1cm} \epsilon^{ij}B_j=\pm B^i,
\end{equation}
In this case there isn't any relation between $A^i$ and $B^i$, because the conditions \eqref{bbcc} decouple the boundary equations \eqref{be31} and \eqref{be32}.
\end{enumerate}

\subsection{Gauge fixing and residual gauge invariance}

In order to study the dynamics of the model, we need to fix a gauge and couple the fields to external sources. We define the total action:
\begin{equation}\label{stot3}
\mathcal S_{tot}= \mathcal S_{bulk} + \mathcal S_{gf} + \mathcal S_J + \mathcal S_{bd},
\end{equation}
where:
\begin{equation}
S_{gf}=\int{d^3x\,\left[bA_2 + dB_2\right]\theta(x^2)}
\end{equation}
fixes the axial gauge choice with the introduction of the Lagrange multipliers $b$ and $d$ ,which corresponds to the gauge condition:
\begin{equation}
A_2=B_2=0,
\end{equation}
while:
\begin{equation}
\mathcal S_J=\int{d^3x\, \left[J^i A_i + K^i B_i\right]\theta(x^2)}
\end{equation}
couples the gauge fields $A$ and $B$ to auxiliary external sources $J$ and $K$ respectively. The action \eqref{stot3} is still invariant under gauge transformations that do not depend on $x^2$ and the residual gauge invariance is functionally expressed by two Ward identities (one for each symmetry $\delta^{(1)}$ and $\delta^{(2)}$):
\begin{gather}
\left[\partial_i J^i + \partial_2 b \right]\theta(x^2)=0
\\
\left[\partial_i K^i + \partial_2 d \right]\theta(x^2)=0.
\end{gather}
Integrating over the $x^2$ coordinate and using the equations of motions derived from \eqref{stot3} (which now include also the Lagrange multipliers $b$ and $d$), the Ward identities become:
\begin{gather}
\int_{0}^{\infty}{dx^2\, \partial_i J^i}= \left.(k-1)\epsilon^{ij}\partial_i B_j\right|_{x^2=0}\label{wi31}
\\
\int_{0}^{\infty}{dx^2\, \partial_i K^i}= \left.(k-1)\epsilon^{ij}\partial_i A_j\right|_{x^2=0}.\label{wi32}
\end{gather}
Going on shell ({\it i.e.} putting $J=K=0$) we recover the conditions \eqref{gic31} and \eqref{gic32}.

\subsection{Boundary algebra and 2D boundary action}

From the conditions of gauge invariance \eqref{gic31} and \eqref{gic32}, it is possible to identify the gauge fields in terms of derivatives of two scalar fields, $\Lambda$ and $\zeta$:
\begin{gather}
\left.\epsilon^{ij}\partial_i B_j\right|_{x^2=0}=0 \Rightarrow \left.B_i\right|_{x^2=0}=\partial_i \zeta(X)\label{qwe}
\\
\left.\epsilon^{ij}\partial_i A_j \right|_{x^2=0}=0 \Rightarrow \left.A_i\right|_{x^2=0}=\partial_i \Lambda(X).
\end{gather}
Their canonical mass dimensions are:
\begin{equation}
\left[\zeta\right]=\left[\Lambda\right]=0
\end{equation}
and their definitions induce the {shift} symmetries:
\begin{gather}
\delta\zeta=c\label{ghi}
\\
\delta \Lambda=c',\label{ghj}
\end{gather}
with $c,c'$ constants. Deriving the Ward identity \eqref{wi31} with respect to $J^i(x')$ one obtains the following equal time commutation relation:
\begin{equation}
\left.(1-k)\left[B_1 (X), A_{1} (X')\right]\right|_{x^0=x'^0}=i\partial_{1}\delta(x^1-x'^1).
\end{equation}
And with similar differentiations:
\begin{gather}
\left.\left[A_1 (X), A_{j} (X')\right]\right|_{x^0=x'^0}=0
\\
\left.\left[B_1 (X), B_{j} (X')\right]\right|_{x^0=x'^0}=0.\label{cmb}
\end{gather}
The commutation relations, written in terms of the scalar fields $\Lambda$ and $\zeta$, take the form:
\begin{gather}
\left.(1-k)\left[\zeta (X),\partial'_1 \Lambda(X') \right]\right|_{x^0=x'^0}=i\delta(x^1-x'^1)\label{comm31}
\\
\left.\left[\zeta (X), \zeta (X') \right]\right|_{x^0=x'^0}=0\label{cmz}
\\
\left.\left[\partial_1\Lambda (X),\partial'_1 \Lambda (X') \right]\right|_{x^0=x'^0}=0.\label{cml}
\end{gather}
Notice that the form of \eqref{cmb} would imply that the commutation relation in \eqref{cmz} is a c-number, but it must be 0 since the commutation relation needs to change {\rm sign} under the exchange $X \leftrightarrow X'$. The relations \eqref{comm31},\eqref{cmz} and \eqref{cml} can be interpreted as canonical commutation relations between the conjugate variables:
\begin{gather}
q(X)\equiv(1-k)\zeta (X)\label{q3}
\\
p(X) \equiv\partial_1 \Lambda (X)\label{p3}.
\end{gather}
The final task of this section is to construct a 2D boundary action which is {invariant under \eqref{ghi} and \eqref{ghj}}, and compatible with the boundary conditions and the definition of the canonical variables \eqref{q3} and \eqref{p3}.
The kinetic term of the corresponding Lagrangian will be:
\begin{equation}
\mathcal L_{kin}=p \dot q =(1-k)\partial_0\zeta\partial_1\Lambda.
\end{equation}
The potential terms must be invariant under \eqref{ghi} and \eqref{ghj} and cannot contain time derivatives. The most general gauge invariant action is:
\begin{equation}
\mathcal S=\int{d^2X\,\left[(1-k)\partial_0\zeta\partial_1\Lambda + \frac{c_1}{2} (\partial_1 \zeta)^2 + \frac{c_2}{2} (\partial_1 \Lambda)^2 + c_3 \partial_1 \zeta \partial_1 \Lambda\right]},
\end{equation}
where $c_1, c_2, c_3$ are constants to be determined. Its equations of motions are:
\begin{gather}
\frac{\delta \mathcal S}{\delta \zeta}=(k-1)\partial_0\partial_1\Lambda -c_1 \partial_1^2\zeta -c_3\partial_1^2\Lambda=0
\\
\frac{\delta \mathcal S}{\delta \Lambda}=(k-1)\partial_0\partial_1\zeta-c_2\partial_1^2\Lambda -c_3\partial_1^2\zeta=0,
\end{gather}
which can be written as:
\begin{gather}
\partial_1\left[(1-k)\partial_0\Lambda +c_1 \partial_1\zeta +c_3\partial_1\Lambda\right]=0
\\
\partial_1\left[(1-k)\partial_0\zeta+c_2\partial_1\Lambda +c_3\partial_1\zeta\right]=0.
\end{gather}
The equations of motions must be compatible with the solutions of the boundary equations discussed in section \ref{boco}. We study each solution separately:
\begin{enumerate}
\item Theory invariant under Time Reversal also on the boundary. The boundary condition \eqref{bcf3} can be {written in terms of} the fields $\zeta$ and $\Lambda$:
\begin{equation}\label{dual3}
a_1 \partial^i \Lambda-\epsilon^{ij}\partial_j\zeta=0.
\end{equation}

The compatibility between \eqref{dual3} and the equations of motion fixes the values of the constants:
\begin{gather}
c_1=\frac{1-k}{a_1}
\\
c_2=a_1(1-k)
\\
c_3=0.
\end{gather}
The 2D boundary action takes the following form:
\begin{equation}\label{s2d}
\mathcal S^{(1)}_{2D}=(1-k)\int{d^2X\, \left[\partial_{0} \zeta\partial_1\Lambda +\frac{1}{2a_1} (\partial_1 \zeta)^2 +\frac{a_1}{2} (\partial_1 \Lambda)^2 \right]}.
\end{equation}
It corresponds to the theory of the Luttinger liquid \cite{Tomonaga:1950zz,Luttinger:1963zz}. Notice that the positivity of the Hamiltonian density associated to the action \eqref{s2d} imply that {$a_1(k-1)>0$}. We remark that the action is left invariant by the exchange of the two fields, provided that the coupling constant $a_1$ goes into its reciprocal:
\begin{gather}
\zeta \leftrightarrow \Lambda
\\
a_1 \rightarrow \frac{1}{a_1}.
\end{gather}
This is a strong-weak coupling duality, which in our case emerges naturally as a consequence of the bulk gauge symmetry.
\\
The action \eqref{s2d} is written in a non covariant way, but it is possible to verify its covariance by means of a criterion proposed by Schwinger \cite{Schwinger:1963zza}, which concerns the algebra formed by  the components of the stress-energy tensor:
\begin{equation}\label{schw}
i\left[T^{00}(X),T^{00}(X')\right]=\left[T^{0\alpha}(X) + T^{0\alpha}(X')\right]\partial_{\alpha} \delta(X-X').
\end{equation}
We compute explicitly the components of the stress-energy tensor:
\begin{gather}
T^{00}=\frac{k-1}{2}\left[\frac{1}{a_1}(\partial_1 \zeta)^2 + a_1 (\partial_1 \Lambda)^2\right]
\\
T^{01}=(k-1)\partial_1 \zeta \partial_1 \Lambda,
\end{gather}
and verify the identity \eqref{schw}:
\begin{multline}
i\left[T^{00}(X),T^{00}(X')\right]=
\\
i\frac{(1-k)^2}{4}\left\{\left[(\partial_1 \zeta(X))^2,(\partial'_1 \Lambda(X'))^2\right]+ \left[(\partial_1 \Lambda(X))^2,(\partial'_1 \zeta(X'))^2\right]\right\}=
\\
-(1-k)\left[\partial_1 \zeta(X) \partial'_1 \Lambda (X') + \partial_1 \Lambda(X) \partial'_1 \zeta (X')\right]\partial_1 \delta(x^1-x'^1)=
\\
\left[T^{01}(X) + T^{01}(X')\right]\partial_1 \delta(x^1-x'^1),
\end{multline}
where we have used the commutation relations \eqref{comm31}, {\eqref{cmz} and \eqref{cml}}.
\\
The covariance can also be realized on-shell. In fact, eliminating the field $\Lambda$ through \eqref{dual3}, the action \eqref{s2d} becomes the one of a free massless scalar field:
\begin{equation}
\mathcal S^{(1)}_{2D}=\frac{(1-k)}{2a_1}\int{d^2X\,\partial_i \zeta \partial^i \zeta}.
\end{equation}
Eliminating $\zeta$ the action \eqref{s2d} becomes:
\begin{equation}
\mathcal S^{(1)}_{2D}=\frac{a_1(1-k)}{2}\int{d^2X\,\partial_i \Lambda \partial^i \Lambda},
\end{equation}
confirming the symmetry under the exchange of the two fields and inversion of the constant $a_1$. The fact that $\Lambda$ and $\zeta$ are free scalars can be also obtained simply as a consequence of \eqref{dual3}, applying a derivative $\partial_i$ or $\partial_j$ on it.

\item The Time Reversal invariance is broken on the boundary. The boundary conditions \eqref{bc333} become:
\begin{gather}
\epsilon^{ij}\partial_j \Lambda=\pm \partial_i \Lambda
\\
\epsilon^{ij}\partial_j \zeta=\pm \partial_i \zeta.
\end{gather}
They are compatible with the equations of motion if:
\begin{gather}
c_1=c_2=0
\\
c_3=\pm(1-k)=\pm 2.
\end{gather}
{
The corresponding 2D boundary action is
\begin{equation}
\mathcal S^{(2)}_{2D}=2\int d^2X(\partial_0\zeta\pm\partial_1\zeta)\partial_1\Lambda, 
\end{equation} 
whose corresponding Hamiltonian density is
\begin{equation}
T^{00}=\mp 2\partial_1\zeta\partial_1\Lambda.
\label{nogood}\end{equation}
Since \eqref{nogood} is not definite positive, we must discard also this solution, leaving \eqref{s2d} as the unique 2D boundary action from the 3D topological BF theory.}
\end{enumerate}


\section{Generalization to {BF models in} higher dimensions}\label{secbf45}

In this section we generalize the method presented in section \ref{secbf3}, applying it to the Abelian BF models in higher dimensions. In particular, we treat the Abelian 4D and 5D cases. The procedure is  analogous to the one used in the previous section, so we skip most of the calculations, stressing our attention on the solutions of the boundary conditions and on the $D-1$-dimensional boundary action. {For $D > 3$ the Time Reversal symmetry doesn't select any term since the Time Reversal invariance is always preserved both in the bulk and on the boundary.}

\subsection{The Abelian 4D BF model with boundary}\label{secbf4}

The action of the Abelian 4D BF model with a boundary on the plane $x^3=0$, is:
\begin{equation}\label{sbf4}
\mathcal S_{bulk}=\int{d^4x\,\epsilon^{\mu\nu\rho\sigma}\left[\partial_{\mu}A_{\nu}B_{\rho\sigma} + k A_{\nu}\partial_{\mu}B_{\rho\sigma} \right] \theta(x^3)},
\end{equation}
with $k \ne 1$. It depends on the gauge field $A_{\mu}$ and on the rank-2 tensor field $B_{\mu\nu}$. {The condition on the canonical mass dimensions is} $\left[A\right] +\left[B\right]=3$. We make the choice $\left[A\right]=\left[B\right]=\frac{3}{2}$ {and we choose the axial gauge  $A_3=B_{3i}=0$}. The most general boundary action compatible with locality and power counting is:
\begin{equation}\label{sbd4}
\mathcal S_{bd}=\int{d^4x\,\left[ \frac{a_1}{2} A^i A_i + \frac{a_2}{2} B^{ij}B_{ij} \right] \delta(x^3)},
\end{equation}
where $a_1, a_2$ are constant parameters.
The boundary conditions and the gauge invariance requirement define the fields on the boundary:
\begin{gather}
\left.\epsilon^{ijk}\partial_i  B_{jk}\right|_{x^3=0}=0 \Rightarrow  B_{ij}=\partial_i \zeta_j(X)-\partial_j \zeta_i(X)\label{qwe4}
\\
\left.\epsilon^{ijk}\partial_j A_k\right|_{x^3=0}=0 \Rightarrow A_i=\partial_i \Lambda(X).
\end{gather}
The definitions of the scalar field $\Lambda$ and the vector field $\zeta_i$ induces the gauge invariance for the vector field $\zeta_i$ and the translation invariance for the scalar field $\Lambda$:
\begin{gather}
\delta \zeta_i=\partial_i \theta
\\
\delta \Lambda= c
\end{gather}
and their canonical mass dimensions are $\left[\Lambda\right]=\left[\zeta\right]=\frac 12$. The boundary conditions of the model can be reduced to a unique one:
\begin{equation}\label{bc4}
\left.\epsilon^{ijk}B_{jk} + a_1 A^i\right|_{x^3=0}=0,
\end{equation}
which, written as a relations between the 3D boundary fields, becomes:
\begin{equation}\label{dual4}
 a_1 \partial^i \Lambda + 2\epsilon^{ijk}\partial_{j}\zeta_k =0.
\end{equation}
with the condition {between the parameters appearing in \eqref{sbd4}:}
\begin{equation}
a_1 a_2=-2k.
\end{equation}
{From the Ward identities describing the residual gauge invariance on the boundary, we get the relevant commutation relation between the boundary fields:
\begin{equation}\label{cmm4dd}
\left.2(k-1)\epsilon^{\alpha\beta} \left[\Lambda (X), \partial'_{\alpha}\zeta_{\beta} (X')\right]\right|_{x^0=x'^0}=i\delta^{(2)}(X-X'),
\end{equation}
while the other ones are simply generalization of \eqref{cmz} and \eqref{cml}.}
\\
The most general gauge and translations invariant 3D action compatible with the boundary condition \eqref{dual4} {and with the commutation relation \eqref{cmm4dd}}, is:
\begin{equation}\label{s3d}
\mathcal S_{3D}=2(k-1)\int{d^3X\,\left[\partial_0\Lambda\epsilon^{0\alpha\beta}\partial_{\alpha} \zeta_{\beta} -\frac{1}{2a_1}F_{\alpha\beta}F^{\alpha\beta} -\frac{a_1}{4}\partial_{\alpha} \Lambda\partial^{\alpha}\Lambda\right]},
\end{equation}
with $F_{\alpha\beta}\equiv\partial_{\alpha}\zeta_{\beta}-\partial_{\beta}\zeta_{\alpha}$ {and where the gauge choice $\zeta_0=0$ has been imposed. Notice that the action \eqref{s3d}} is completely equivalent to the one proposed by \cite{Cho:2012tx} for the study of topological insulators. The relation $a_1(k-1)>0$ holds again. 
{The covariance of the action \eqref{s3d} can be easily checked by means of the Schwinger's criterion \cite{Schwinger:1963zza} on the components of the stress energy tensor $T^{\mu\nu}$, which is satisfied thanks to the crucial commutation relation \eqref{cmm4dd}. Alternatively, }
%
we can show the covariance on-shell with the elimination of the field $\Lambda$ through \eqref{dual4}. The action \eqref{s3d} becomes:
\begin{equation}
\mathcal S_{3D}=\frac{(1-k)}{a_1}\int  d^3X F_{ij}F^{ij},
\end{equation}
where the gauge condition $\zeta_0=0$ has not been imposed. Remarkably, we obtain the 3D Maxwell theory on the boundary of the 4D topological BF model. In the same way, eliminating the field $\zeta_i$, we obtain the action of a free scalar, analogously to the previous model:
\begin{equation}
\mathcal S_{3D}=\frac{a_1(1-k)}{2}\int  d^3X\partial_i\Lambda \partial^i \Lambda.
\end{equation}


\subsection{The Abelian 5D BF model with boundary}\label{secbf5}

The action of the Abelian 5D BF model with a boundary on the plane $x^4=0$, is:
\begin{equation}\label{sbf5}
\mathcal S_{bulk}=\int{d^5x\,\epsilon^{\mu\nu\rho\sigma\tau}\left[\partial_{\mu}A_{\nu}B_{\rho\sigma\tau} + k A_{\nu}\partial_{\mu}B_{\rho\sigma\tau} \right] \theta(x^4)},
\end{equation}
with $k \ne 1$. It depends on the gauge field $A_{\mu}$ and the rank-3 tensor $B_{\mu\nu\rho}$. We as{\rm sign} to the fields the canonical mass dimensions $\left[A\right]=\left[B\right]=2$ {and we make the usual axial gauge choice $A_4=B_{4ij}=0$}. The boundary term is:
\begin{equation}
\mathcal S_{bd}=\int{d^5x\, \left[\frac{a_1}{2} A^i A_i + \frac{a_2}{2} B^{ijk} B_{ijk} + a_3 \epsilon^{ijkl}B_{ijm}B^{m}_{kl}  \right]\delta(x^4)}.\label{sbd5}
\end{equation}
The gauge invariance identify the boundary fields as a scalar $\Lambda$ and an {antisymmetric} rank-2 tensor $\zeta_{ij}$:
\begin{gather}
\left.\epsilon^{ijkl}\partial_i B_{jkl}\right|_{x^4=0}=0 \Rightarrow B_{ijk}=\partial_{i} \zeta_{jk}(X) + \mbox{cyclic permutations}\label{qwe5}
\\
\left.\epsilon^{ijkl}\partial_k A_l\right|_{x^4=0}=0 \Rightarrow A_i=\partial_i \Lambda(X),
\end{gather}
{together with the symmetries}:
\begin{gather}
\delta \zeta_{ij}=\partial_i \theta_j - \partial_j \theta_i
\\
\delta \Lambda= c.
\end{gather}
{The boundary condition is}:
\begin{equation}\label{bc5}
\left.\epsilon^{ijkl}B_{jkl}+ a_1 A^i\right|_{x^4=0}=0,
\end{equation}
{where the constant parameters appearing in \eqref{sbd5} are constrained as follows}:
\begin{equation}
a_3=0, \hspace{1cm} a_1a_2=-6k,
\end{equation}
The boundary condition \eqref{bc5} written in terms of $\Lambda$ and $\zeta_{\alpha\beta}$ is:
\begin{equation}\label{dual5}
3\epsilon^{ijkl}\partial_j \zeta_{kl}+ a_1 \partial^i\Lambda=0.
\end{equation}
{In close analogy with the previous analysis, the commutation relation between the boundary fields is:
\begin{equation}\label{cmm5dd}
\left.3(1-k)\epsilon^{\alpha\beta\gamma}\left[\Lambda (X),\partial'_{\alpha} \zeta_{\beta\gamma}(X') \right]\right|_{x^0=x'^0}=i\delta^{(3)}(X-X').
\end{equation}}
Finally, the gauge invariant 4D boundary action compatible with \eqref{dual5} is:
\begin{equation}\label{s4d}
\mathcal S_{4D}=3(1-k)\int{d^4X\,\left[\partial_0\Lambda\epsilon^{0\alpha\beta\gamma}\partial_{\alpha} \zeta_{\beta\gamma} + \frac{3}{2a_1}(\epsilon^{0\alpha\beta\gamma}\partial_{\alpha} \zeta_{\beta\gamma})^2 + \frac{a_1}{6} \partial_{\alpha} \Lambda \partial^{\alpha}\Lambda\right]},
\end{equation}
with the gauge condition $\zeta_{0\alpha}=0$ and $a_1(k-1)>0$. {The commutation relation \eqref{cmm5dd} between the components of the stress-energy tensor guarantees again the validity of the Schwinger's criterion \eqref{schw}.}
\\
Eliminating the field $\Lambda$ through \eqref{dual5}, the action \eqref{s4d} takes the on-shell covariant form:
\begin{equation}\label{xxx}
\mathcal S_{4D}=\frac{3(1-k)}{a_1}\int d^4X\, F_{ijk}F^{ijk},
\end{equation}
where $F_{ijk}=\partial_i\zeta_{jk} + \partial_k\zeta_{ij} + \partial_j\zeta_{ki}$. Once again, eliminating the filed $\zeta_{ij}$, the action \eqref{s4d} becomes:
\begin{equation}\label{xxy}
\mathcal S_{4D}=\frac{a_1(1-k)}{2}\int d^4X\,\partial_i \Lambda \partial^i \Lambda,
\end{equation}
which confirms the duality of the models with actions \eqref{xxx} and \eqref{xxy}, {as claimed in \cite{Kalb:1974yc}}.


\section{The Abelian 5D BC model with boundary}\label{secbc}

In this section we extend our treatment to the so-called BC model, built from two rank-2 tensors \cite{Kravec:2013pua}, with boundary on the plane $x^4=0$, which was studied in euclidean space-time in \cite{Amoretti:2014kba}. In Minkowski 5D flat space-time its action is defined by :
\begin{equation}\label{sbc}
\mathcal S_{bulk}=\int{d^5x\,\epsilon^{\mu\nu\rho\sigma\tau}\left[\partial_{\rho}B_{\mu\nu}C_{\sigma\tau} + k B_{\mu\nu}\partial_{\rho}C_{\sigma\tau} \right] \theta(x^4)},
\end{equation}
with $k \ne 1$.
{The canonical mass of the tensors $B_{\mu\nu}$ and $C_{\mu\nu}$ are $\left[A\right]=\left[B\right]=2.$} The most general boundary term which can be introduced is:
\begin{equation}
\mathcal S_{bd}=\int{d^5x\, \left[a_1 B^{ij}B_{ij} + a_2 \epsilon^{ijkl}B_{ij}B_{kl} + a_3 C^{ij}C_{ij} + a_4 \epsilon^{ijkl}C_{ij} C_{kl} + a_5 B^{ij}C_{ij} \right]\delta(x^4)}.\label{sbdbc}
\end{equation}
The bulk action \eqref{sbc} is invariant under two discrete symmetries involving {inversion of time}:
\begin{align}
&T_1x^0=-x^0
&
&T_1x^{\alpha,4}=x^{\alpha,4}\notag
\\
&T_1B_{0\alpha}=B_{0\alpha}
&
&T_1C_{0\alpha}=-C_{0\alpha}\notag
\\
&T_1B_{04}=B_{04}
&
&T_1C_{04}=-C_{04}
\\
&T_1B_{\alpha\beta}=-B_{\alpha\beta}\notag
&
&T_1C_{\alpha\beta}=C_{\alpha\beta}
\\
&T_1B_{\alpha4}=-B_{0\alpha4}\notag
&
&T_1C_{\alpha4}=C_{\alpha4}
\end{align}
and:
\begin{align}
&T_2x^0=-x^0
&
&T_2x^{\alpha,4}=x^{\alpha,4}\notag
\\
&T_2B_{0\alpha}=-C_{0\alpha}
&
&T_2C_{0\alpha}=-B_{0\alpha}\notag
\\
&T_2B_{04}=-C_{04}
&
&T_2C_{04}=-B_{04}
\\
&T_2B_{\alpha\beta}=C_{\alpha\beta}\notag
&
&T_2C_{\alpha\beta}=B_{\alpha\beta}
\\
&T_2B_{\alpha4}=C_{0\alpha4}\notag
&
&T_2C_{\alpha4}=B_{\alpha4}.
\end{align}
Notice that $T_2$ is a symmetry only if  $k=-1$. The {vector boundary fields are defined in the usual way from the Ward identities describing the residual gauge invariance on $x^4=0$}:
\begin{gather}
\left.\epsilon^{ijkl}\partial_j C_{kl}\right|_{x^4=0}=0 \Rightarrow \left.C_{ij}\right|_{x^4=0}=\partial_i \xi_j(X)-\partial_j \xi_i(X)\label{qwebc}
\\
\left.\epsilon^{ijkl}\partial_j B_{kl}\right|_{x^4=0}=0 \Rightarrow \left.B_{ij}\right|_{x^4=0}=\partial_i \zeta_j(X)-\partial_j \zeta_i(X).
\end{gather}
Their canonical mass dimensions are $\left[\Lambda\right]=\left[\zeta\right]=1$ and their definitions induce the {gauge} symmetries:
\begin{gather}
\delta \xi_i=\partial_i \varphi
\\
\delta \zeta_{i}=\partial_i \theta
\end{gather}
The boundary conditions of the model are:
\begin{gather}
\left.-\epsilon^{ijkl}C_{kl} + 2 a_1 B^{ij} +2a_2 \epsilon^{ijkl}B_{kl} + a_5 C^{ij}\right|_{x^4=0}=0\label{bbcc1}
\\
\left.-k\epsilon^{ijkl}B_{kl} + 2 a_3 C^{ij} + 2 a_4 \epsilon^{ijkl}C_{kl} + a_5 B^{ij} \right|_{x^4=0}=0.\label{bbcc2}
\end{gather}
The only consistent solutions are those which respect Time Reversal also on the boundary \cite{Amoretti:2014kba}. We study them separately:
\begin{enumerate}
\item The solution imposing $T_1$:
\begin{equation}
a_2=a_4=a_5=0, \hspace{1cm} a_1 a_3 =- k,
\end{equation}
with the unique boundary condition:
\begin{equation}\label{bcbc}
\left.-\epsilon^{ijkl}C_{kl} + 2 a_1 B^{ij} \right|_{x^4=0}=0,
\end{equation}
which, written in terms of $\xi_{i}$ and $\zeta_{i}$, is:
\begin{equation}\label{dualbc}
-\epsilon^{ijkl}\partial_k \xi_l +  a_1 \left(\partial^i \zeta^j - \partial^j \zeta^i \right)=0.
\end{equation}
It induces the 4D gauge invariant action:
\begin{equation}\label{s4dbc}
\mathcal S^{(1)}_{4D}=4(1-k)\int{d^4X\,\left[\epsilon^{\alpha\beta\gamma}\partial_0 \xi_{\alpha}\partial_{\beta}\zeta_{\gamma} -\frac{1}{4}\left(\frac{1}{a_1} F_{\alpha\beta}F^{\alpha\beta} + a_1 G_{\alpha\beta}G^{\alpha\beta}\right) \right]}.
\end{equation}
where $F_{\alpha\beta} \equiv \partial_{\alpha}\xi_{\beta}-\partial_{\beta}\xi_{\alpha}$ and $G_{\alpha\beta} \equiv \partial_{\alpha}\zeta_{\beta}-\partial_{\beta}\zeta_{\alpha}$ and and with the gauge choice $\xi_0=\zeta_{0}=0$ and the condition $a_1(k-1)>0$.
The action displays an electromagnetic-like duality, as it is invariant under the symmetry:
\begin{equation}\label{elsim}
\begin{split}
\zeta &\leftrightarrow \xi
\\
a_1 &\rightarrow \frac{1}{a_1}
\end{split}
\end{equation}
which exchanges the ``electric-like'' and ``magnetic-like'' fields.
\\
{Also in this last case, it can be verified that the components of the stress-energy tensor satisfy the Schwinger's identity \eqref{schw}, in virtue of the commutation relation:
\begin{equation}
\left.4(1-k)\epsilon^{\alpha\beta\gamma}\left[ \xi_{\alpha}(X),\partial'_{\beta}\zeta_{\gamma} (X') \right]\right|_{x^0=x'^0}=i\delta^{(3)}(X-X'),
\end{equation}
thus assuring the covariance of the action \eqref{s4dbc}. 
Alternatively, it is possible to check the covariance of the 4D action eliminating the field $\zeta$ through the duality relation \eqref{dualbc}. Remarkably, the resulting action turns out to coincide with the 4D Maxwell theory:}
\begin{equation}\label{lkjh}
\mathcal S^{(1)}_{4D}=\frac{(k-1)}{a_1}\int{d^4X\, F_{ij}F^{ij}}.
\end{equation}
Eliminating $\xi$ we would obtain again the Maxwell theory but with $\xi \rightarrow \zeta$ compared to \eqref{lkjh} and the coupling constant $a_1(k-1)$, in accordance with the {electromagnetic-like duality} \eqref{elsim}.

\item The solution imposing $T_2$:
\begin{equation}
a_3=a_1, \hspace{1cm}  a_4=-a_2, \hspace{1cm} k=-1,
\end{equation}
with the boundary conditions:
\begin{gather}
B^{ij}=\kappa_1 \epsilon^{ijkl}B_{kl} + \kappa_2 \epsilon^{ijkl}C_{kl},\label{bici1}
\\
C^{ij}=-\kappa_2 \epsilon^{ijkl}B_{kl} - \kappa_1 \epsilon^{ijkl}C_{kl},\label{bici2}
\end{gather}
where:
\begin{equation}
\kappa_1^2-\kappa_2^2=-\frac 14,
\end{equation}
and:
\begin{equation}
a_1=4\kappa_1a_2 + 2\kappa_2, \hspace{1cm} a_5=8\kappa_2 a_2 + 4\kappa_1.
\end{equation}
The boundary action takes the following form:
\begin{multline}
\mathcal S^{(2)}_{4D}=8\int{d^4X\,\left[\epsilon^{\alpha\beta\gamma}\partial_0 \xi_{\alpha}\partial_{\beta}\zeta_{\gamma} -\frac 12\kappa_2 F_{\alpha\beta}F^{\alpha\beta} \right.}
\\
{\left.-\frac 12 \kappa_2 G_{\alpha\beta}G^{\alpha\beta} -\kappa_1 F_{\alpha\beta}G^{\alpha\beta}\right]}.
\end{multline}
with the same {temporal} gauge choice of \eqref{s4dbc}. Notice that $\kappa_2$ must be necessarily positive. Otherwise, the respective Hamiltonian would not be definite positive. 

\item The solution imposing $T_1$ and $T_2$ together. It is a special case of 2. with the further conditions:
 \begin{equation}
\kappa_1=0 \hspace{1cm} \kappa_2>0.
\end{equation}
The boundary conditions become:
\begin{gather}
B^{ij}= \frac{1}{2} \epsilon^{ijkl}C_{kl},\label{bici12}
\\
C^{ij}=-\frac{1}{2} \epsilon^{ijkl}B_{kl},\label{bici22}
\end{gather}
which are consistent with each other, and the boundary action is:
\begin{equation}\label{s4dbc21}
\mathcal S^{(3)}_{4D}=8\int{d^4X\,\left[\epsilon^{\alpha\beta\gamma}\partial_0 \xi_{\alpha}\partial_{\beta}\zeta_{\gamma}  - \frac {1}{4}\left( F_{\alpha\beta}F^{\alpha\beta} +  G_{\alpha\beta}G^{\alpha\beta}\right) \right]}.
\end{equation}
It is a special case of \eqref{s4dbc} in the limit $k=-1$ and $a_1=1$ and with the same argument it is straightforward to verify its covariance.
\end{enumerate}

\section{Conclusions}\label{seccon}

In this paper we discuss, in a common framework, some of the topological quantum field theories that we have studied in our previous works, in the presence of a boundary, introduced by means of a theta term in the action. We have been able to identify the boundary physics emerging from bulk theories which otherwise lack of local observables: in all cases we analyzed, gauge symmetry play a crucial role determining which are the boundary fields and the transformations under which the boundary actions must be invariant. The bulk contribution to the boundary actions are therefore uniquely determined by requiring compatibility with the algebra arising from the Ward identities and with the boundary conditions. We obtained that the 2D theory of Luttinger liquid emerges as boundary theory of the 3D BF theory. For higher dimensions, the Maxwell theory is naturally found on the boundary of topological field theories. {We stress that the boundary actions depend on the coefficient $a_i$ appearing in the $\theta$ terms of the various bulk actions we considered. These coefficients are not entirely determined by the symmetries of the bulk theory, as it should, since they encode non-universal information. In addition, for what concerns the $BC$ model studied in section 4, two possible boundary dynamics are found, which reflect the two possible time reversal symmetries displayed by this model.} Moreover, and remarkably, some of the actions displays a strong-weak coupling duality, such as the case for Luttinger theory in 2D and for Maxwell theory in 4D. Finally, despite appearances, we showed that the boundary actions display the Schwinger criterion for covariance, based on algebraic considerations on the energy-momentum tensor.

\newpage

{\bf Acknowledgements}

We thank the support of INFN Scientific Initiative SFT: ``Statistical Field Theory, Low-Dimensional Systems, Integrable Models and Applications'' and FIRB - ``Futuro in Ricerca 2012'' - Project HybridNanoDev RBFR1236VV.


\end{document}